\documentclass{article}
\usepackage{amsmath}

\newcommand{\nocomma}{}
\newcommand{\tmop}[1]{\ensuremath{\operatorname{#1}}}
\newcommand{\tmstrong}[1]{\textbf{#1}}

\begin{document}

\title{Superluminal neutrinos and the Standard Model}\author{\\
Jorge Alfaro\\
Facultad de F\'{\i}sica, Pontificia Universidad Cat\'olica de Chile\\
Casilla 306, Santiago 22, Chile.\\
jalfaro@puc.cl}
\maketitle

\begin{abstract}
Recently the OPERA collaboration {\cite{opera}} has reported the
observation of superluminal neutrinos traveling a distance of 730 km from
Grand Sasso Laboratory to CERN.These results contradict the basic tenet of
the Theory of Special Relativity: No particle can travel faster than light in
vacuum. Moreover \ they \ seem to be in conflict with the speed of
(anti)neutrinos detected from the explosion of the SP1987A Super
Nova{\cite{SP1987A}}.Here we show that the relative velocity between neutrino
and photon has the following property: It depends weakly on the energy of the
particle excepts in certain regions(thresholds) where discrete jumps
appear.This explains both Opera and SP1987A Super Nova data.
\end{abstract}

OPERA measurement of the relative velocity between neutrino and photon(RVN) is
\begin{equation}
  \frac{v_{\nu} - v_{\gamma}}{v_{\gamma}} = 2.48 \pm 0.28 \left( \tmop{stat} .
  \right) \pm 0.30 \left( \tmop{sys} . \right) \times 10^{- 5} \label{opera}
\end{equation}
In the past the \ MINOS experiment{\cite{minos}} reported a RVN of
\[ \frac{v_{\nu} - v_{\gamma}}{v_{\gamma}} = 5.1 \pm 2.9 \times 10^{- 5} \]

If we accept (\ref{opera}), the neutrinos from the Super Nova should have
arrived years before light reaches us from the explosion. But at most a retard
of few hours was observed.

Therefore the \ speed of (anti)neutrinos detected from the explosion of the
SP1987A must satisfy:

\[ \frac{\left| v_{\nu} - v_{\gamma} \right|}{v_{\gamma}} < 2 \times 10^{- 9}
\]

In all experiments the speed of the neutrinos  do not show a dependence on the
energy of the particle.

OPERA deals with $\mu$ neutrinos at an energy of 20-40 Gev. The Super Nova
Neutrinos are mainly $e \nocomma, \mu$ antineutrinos with an energy of 10-20
Mev. That is OPERA neutrinos have 1000 times more energy than the Supernova
neutrinos.

In {\cite{prl}}, we proposed a mechanism of Lorentz Invariance Violation(LIV)
that within the Standard Model of Particle Physics, predicted that the
particles will have different Maximal Attainable Velocities(MAV) {\cite{cg}}.
In particular neutrinos, photons and electrons will travel at different
maximal speeds, because their mutual interactions are different.

For alternative scenarios to break Lorentz invariance, see {\cite{otros}}; or
deform it see {\cite{dsr}}.

In this letter, we want to show how the results of {\cite{prl}} can be used to
explain the observations of OPERA and SP1987A Super Nova and obtain new
predictions to be tested in future experiments. We will see that the RVN has
the following property: It depends weakly on the energy of the particle
excepts in certain regions(thresholds) where discrete jumps appear.

The RVN computed in \ {\cite{prl}} is:

\begin{eqnarray}
  \frac{v_{\nu} - v_{\gamma}}{v_{\gamma}} = e^2 \alpha \left\{ \right.
  \frac{2}{3} \sum_f N_{c f} Q_f^2 - \frac{3}{2} - \frac{(3 + tan^2 \theta_w)}{8 s i
  n^2 \theta_w}  \left\} \right. &  &  \label{liv}
\end{eqnarray}
where $Q_f$ and $N_{c f}$ are the electric charge and the color factor(1 for
leptons, 3 for quarks) of the fermion $f$, $e$ is the coupling of
the photon to a charged particle and $\theta_w$ is the Weinberg's angle. The
factor $\frac{3}{2}$ in the curly bracket is the contribution of the $W_+$
vector boson.$\alpha$ is a free constant that parametrizes the breaking of Lorentz symmetry at
high energies.

In the Standard Model $e, \theta_w$ run with the energy scale. In the range of
energies we are studying: 20-40 Gev for the Opera neutrino and 10-20 Mev for
the Super Nova neutrino, the change of the coupling constants is
small{\cite{PDG}}.

The main variation comes from the threshold in the photon vacuum polarization.
The charged particles to be considered in (\ref{liv}) must have masses
smaller than the neutrino energy \ $E$ {\cite{Pokorski}}.

For energies below $m_{\mu} = 105 \tmop{Mev}$(muon mass), we have to include
the electron and quarks u,d.

For energies $m_b < E < m_W$, where $m_b = 4.2 \tmop{Gev}$(bottom quark mass)and $m_W = 80 \tmop{Gev}$(W-boson mass) we must add the muon,the tau and quarks s,c,b.

In both cases, $W_+$ is not included, because it has a higher mass.

So we get:
\begin{eqnarray}
  \frac{v_{\nu} - v_{\gamma}}{v_{\gamma}} \left( 0 < E < m_{\mu}\right) = e^2 \alpha \left\{ \right. \frac{16}{9} - \frac{(3 +
  tan^2 \theta_w)}{8 s i n^2 \theta_w}  \left\} \right. = e^2 \alpha
  f_{\tmop{LOW}} &  &  \label{low}\\
  \frac{v_{\nu} - v_{\gamma}}{v_{\gamma}} \left( m_{b} < E < m_W\right) = e^2 \alpha \left\{ \right. \frac{40}{9} -
  \frac{(3 + tan^2 \theta_w)}{8 s i n^2 \theta_w}  \left\} \right. = e^2
  \alpha f_{\tmop{HIGH}} &  &  \label{high}
\end{eqnarray}
The main uncertainties come from the different schemes to compute $\theta_w$.

Allowing a 4\% increase in $\theta_w$, we get an estimation of Weinberg angle
at low energy, using Table 10.2 from \ {\cite{PDG}} :
\[ 0.2322 < s i n^2 \theta_w < 0.2409 \nocomma ; - 1.2 \times 10^{- 5} <
   f_{\tmop{LOW}} < 5.6 \times 10^{- 2} \]

We see that the value of $\theta_w$ at low energies is compatible with
$f_{\tmop{LOW}} = 0$.

In fact
\[ f_{\tmop{LOW}} = 0 \tmop{at} s i n^2 \theta_w = 0.232202 \]
From (\ref{opera}) and (\ref{high}), we get $\alpha \sim 10^{- 4}$.

\section{electron MAV}

Define: $e_L = \frac{1 - \gamma^5}{2} e$, $e_R = \frac{1 + \gamma^5}{2} e$,
where $e$ is the electron field. We get \cite{prl}:
\begin{eqnarray}
  c_L = 1 - ( \frac{g^2}{\cos^2 \theta_w} (\sin^2 \theta_w - 1 / 2)^2 + e^2 +
  g^2 / 2) \frac{\alpha}{2} ; \\
  c_R = 1 - (e^2 + \frac{g^2 \sin^4 \theta_w}{\cos^2 \theta_w})
  \frac{\alpha}{2}
\end{eqnarray}

\section{Relative velocity neutrino electron}

\begin{eqnarray}
  c_{\nu} - c_L = 0\\
  c_{\nu} - c_R = \frac{3 e^2 \alpha}{8}
  \left\{ \sec^2 \theta_w - \tmop{cosec}^2 \theta_w \right\} 
\end{eqnarray}

\section{Pair creation}

We want to study the kinematics of the process  $\nu_{\mu} \rightarrow \nu_{\mu}
+ e^+ + e^-$ that has been proposed in \cite{cog} to rule out neutrino superluminar speeds in Opera. Icarus\cite{icarus}
did not observe this decay mode in Opera, thus claiming that there cannot be superluminar neutrinos in Opera. In our model this process is forbidden, so it should not be observed. Icarus result can be interpreted as supporting this view. 
The threshold energy above which the process is allowed is\cite{cog}:
\begin{eqnarray*}
  E_0=\frac{\sqrt{2}m_e}{\sqrt{c_\nu-c_e}}
\end{eqnarray*}
In our model, this process is forbidden. The left electron has $\delta =
c_{\nu_{\mu}} - c_e$=0, whereas the righ electron has $\delta < 0$. So Icarus
evidence is inconclusive.

\section{Compatibility with MINOS\cite{minos}}
In MINOS the velocity of a 3 GeV neutrino beam is measured by comparing detection times
at the near and far detectors of the experiment, separated by 734 km. They measure
$\frac{v - c}{c}
= 5.1 \pm 2.9 \times 10^{- 5}$(at 68\% C.L.).

In our model, the velocity of the neutrinos changes at thresholds where new SM
particles enter the loop, in particular the charm (1.3 GeV), tau (1.77
GeV) and bottom (4.5 GeV). These thresholds fall into the spectrum of
MINOS and they should have been observed. The fact is that the uncertainty in the measurement of the velocity
of the neutrino reported by MINOS is too large to see the velocity jumps.
The maximum change in the velocity after including all the particles mentioned above in (\ref{liv}) will be:
\begin{eqnarray*}
  \Delta v_\nu = \frac{16}{9} e^2 \alpha = 1.66 \times 10^{- 5}
\end{eqnarray*}
It is well into the uncertainty of MINOS.

\section{Compatibility with FERMILAB measurement of the relative neutrino velocity}

In \cite{fermilab} several relative velocities of neutrino, antineutrino and muon were measured with great precision. They reported that no energy dependence of the velocities of neutrinos or antineutrinos is
  observed within the statistical and systematic errors over the energy range
  80 to 200 GeV. The velocity differences (95\% confidence level) are
  $\left| \beta_{\nu} - \beta_{\bar{\nu}} \right| < 0.7 \times 10^{- 4}$,
  $\left| \beta_{\nu_K} - \beta_{\nu_{\pi}} \right| < 0.5 \times 10^{- 4}$,
  $\left| \beta_{\nu \left( \bar{\nu} \right)} - \beta_{\mu} \left(
  \tmop{corrected} \right) \right| < 0.4 \times 10^{- 4}$ where $\beta = v /
  c$. 
  
We have that , in this model, $\beta_{\nu} - \beta_{\bar{\nu}} = 0$,
$\beta_{\nu_K} - \beta_{\nu_{\pi}} = 0$, $\beta_{\nu \left( \bar{\nu} \right)}
- \beta_{\mu_L} = 0$ . It remains to check:
\begin{eqnarray*}
  c_{\nu} - c_R = \frac{3 e^2 \alpha}{8} \left\{ \sec^2 \theta_w -
  \tmop{cosec}^2 \theta_w \right\} = - \frac{9}{8} e^2 \alpha = - 1.04 \times 10^{- 5} &  & 
\end{eqnarray*}
The bound of \cite{fermilab} are satisfied in this model.

The main predictions to be tested in future experiments are:

1)Jumps in the relative velocity when crossing thresholds at the masses of
massive particles in the Standard Model.Corrections ${\cal O }\left( \left( \frac{E}{M} \right)^2 \right)$are expected, where
$M$ is the mass of the lightest decoupled particle\cite{Pokorski}.

2)Slow variation of the relative velocity with energy(running of the
couplings).

3)MAV is independent of the family, i.e. electron, muon and tau neutrinos must have the same MAV.

4) In addition to this, we should observe birefringence in charged leptons, but not in gauge bosons.
The right handed electron(muon,tau) should move at a different maximal speed
compared to the left handed electron(muon,tau)\cite{prl}.

{\tmstrong{Note Added}} For various discussions of the OPERA report, please
\ see{\cite{wopera}}. 

{\tmstrong{Acknowledgements}}. J.A. thanks M. Loewe for a useful discussion and carefully reading the manuscript.
He also want to acknowledge useful comments from G: Cacciapaglia.
The work of JA is partially supported by VRAID/DID/46/2010 and Fondecyt
1110378.

\end{document}